\documentclass[journal,twocolumn,10pt]{IEEEtran} 
\usepackage{epsfig,cite}
\usepackage{graphicx}
\graphicspath{ {./figures/} }
\usepackage{tabularx}
\usepackage{caption}
\usepackage{subcaption}
\usepackage[font=small,labelfont=bf]{caption}
\usepackage[T1]{fontenc}
\usepackage{mathptmx}
\usepackage{color}
\usepackage{amssymb,amsmath,mathrsfs}
\usepackage[left=1.587cm,right=1.824cm,top=1.905cm,bottom=2.54cm]{geometry}
\usepackage{url}

\newcommand{\rev}[1]{\textcolor{black}{#1}}

\begin{document}
\title{Designing a Light-based Communication System with a Biomolecular Receiver}
\author{
	\IEEEauthorblockN{Taha Sajjad and Andrew W. Eckford}%
    \thanks{The authors are with the Department of Electrical Engineering and Computer Science, York University, 4700 Keele Street, Toronto, Ontario, Canada M3J 1P3. Emails:  tahaatiq@cse.yorku.ca, aeckford@yorku.ca}%
	\thanks{This work was supported by a grant from the Lloyd's Register Foundation International Consortium of Nanotechnologies (LRF ICoN).}%
 \thanks{Material in this paper was presented in part at the 10th ACM International Conference on Nanoscale Computing and Communication (NANOCOM), Coventry, UK, September 20-22, 2023.}
}

\maketitle
\begin{abstract}
 Biological systems transduce signals from their surroundings in numerous ways. This paper introduces a communication system using the light-gated ion channel Channelrhodopsin-2 (ChR2), which causes an ion current to flow in response to light. Our design includes a ChR2-based receiver along with encoding, modulation techniques and detection. 
 Analyzing the resulting communication system, we discuss the effect of different parameters on the performance of the system.
 Finally, we  discuss its potential design in the context of bio-engineering and light-based communication and show that the data rate scales up with the number of receptors, indicating that high-speed communication may be possible.
\end{abstract}

\section{Introduction}
Living cells acquire information from their surroundings through numerous signal transduction pathways. For example, specialized cells in multicellular organisms must communicate with each other to coordinate their actions. Signal transduction takes many forms: the input signal can be carried by changes in chemical concentration, electrical potential, light intensity, mechanical forces, and temperature, among others. 

In this paper, we are concerned with light-based communication. 
From an engineering perspective, light communications is a promising technology to provide reliable high-speed communication \cite{sindhubala2015design}, \cite{idris2019visible}, such as through LiFi (Light Fidelity). Additionally, light also serves as a means of communication among living organisms.
In this paper, we develop a communication model based on the light-gated receptor channelrhodopsin-2 (ChR2), taking advantage of its light-sensing properties to design a biological receiver.
Our motivation for using ChR2 is its remarkable receiving capacity, as presented in \cite{ thomas2016shannon}, indicating that the information rate increases linearly with the number of receptors. 
Channelrhodopsins, derived from the green alga and producing photocurrent in response to blue light, could potentially replace silicon photo receivers in the future, allowing biodegradable components to be used in communication systems which could be constructed using techniques from systems biology.

Light-based communication and control is used in a variety of biological and biotechnological contexts. For example ChR2 is a key tool in the emerging technique of optogenetics, used to modulate neuronal activity \cite{optogenetics2016,nagel2003channelrhodopsin,lorenz2014channelrhodopsin}. Optogenetic methods have already begun to pave their way into molecular communication systems \cite{barros2018feed}, such as the optical controller proposed in \cite{krishnaswamy2020shining} to control the molecular communication system at the microscale. This controller triggers information processing and communication that reduces the computational complexity of a molecular communication system. Similarly, in \cite{grebenstein2018biological}, a light-based microscale modulator has been proposed that converts optical signal to chemical signal. There are several methodologies to simulate, control and predict the output of neural networks \cite{luo2016optogenetics}, \cite{noel2018distortion}, which require communication among various components of the neural processor on the chip \cite{guryanov2016receptor}, \cite{javdan2021design}, \cite{noel2019modeling}. 
\rev{A ChR2 based communication system can also be used in conjunction with other systems over the internet-of-every-nano-thing as mentioned in \cite{chouhan2023interfacing}.}  ChR2-based signalling is also closely related to signal transduction.
In \cite{channelrhodopsin2018,thomas2016shannon}, signal transduction was described as a communication system. The Shannon capacity of signal transduction has been calculated for simplified Markov models with slowly changing inputs \cite{einolghozati2011capacity} and for populations of bacteria interacting with each other \cite{aminian2015capacity}. These papers focused on the mutual information and capacity of finite-state signal transduction channels of different biological receptors, including the ligand-gated receptor acetylcholine (ACh) and the light-gated receptor channelrhodopsin-2 (ChR2). The biological receptors are key components of bioengineering with applications in molecular communications and system biology. However, these earlier works did not discuss practical communication system design.\par 

\rev{A framework for communication using biomolecular receiver has been presented in \cite{sajjad2023designing}. In the present work, we extend our results from a single receptor case to a system involving multiple receptors where these receptors are {\em independent and statistically identical.} Additionally, we analyze the impact of noise on our system.} The system is modelled as a discrete-time, finite-state Markov chain with transition rates governed by the input signal\cite{thomas2016shannon}. We  determine the information rate of a system by applying results of information theory established in \cite{channelrhodopsin2018}, \cite{thomas2016shannon}, validated by simulations.\par
The paper is organized as follows. In Section II, we describe our basic system and its theoretical aspects. In Section III, we present a communication model to analyze the performance of our proposed system. In Section IV, we  extended our model by considering multiple receptors and presented our  results in terms of data rate in Section V. In Section VI, we conclude and discuss the prospects in this domain.

\section{Biophysical model}

Here, we introduce a basic communication model including transmitter, channel and receiver. Our model is inspired by \cite{khan2017visible}, where the transmitter has a source encoder, modulator and light source, while the receiver consists of the photodiode (silicon, PIN, avalanche). We have replaced the photoreceiver with a single biological receptor ChR2 to design different communication system parameters as shown in Fig. \ref{fig1}.

\subsection{Channelrhodopsin Machinery}
 The intracellular responses can often be represented as a transition among discrete states. The same principle applies to the light-gated receptor ChR2. In response to blue light, channelrhodopsins open their ion channel, allowing an ion current to pass in the interior of the cell afterwards, a relaxation process brings the receptor back to the initial state \cite{nagel2003channelrhodopsin},\cite{nikolic2009photocycles}. The transition cycle of ChR2 can be illustrated as a three-state model. As shown in Fig. \ref{fig2}a, ChR2 changes from  {\em Closed} $C1$, to {\em Open} $O2$, then to the intermediate {\em Desensitized} state  $D3$ and finally back to the closed state. The {\em Desensitized} state is a condition where the current stops flowing, \rev{and} the input light does not affect the receptor. In the {\em closed state} $C1$, the light triggers the channel to open. \rev{The model assumes that once ChR2 transitions to a new state, it cannot revert immediately to the previous state. For example, the probability of changing {\em directly} from {\em Open} to {\em Closed} state is $0$, and this principle applies to all other state transitions as well, as we will discuss below.}\par

 
 In Fig. \ref{fig2}a, states are represented by a compound label consisting of state property and state number. A solid, thick arrow depicts an input-dependent transition, the light level in this case. State transitions are defined by transition rates, which determine the probability of a state change occurring in an infinitesimally small amount of time. In case of ChR2, the time-dependent light intensity influences the transition from $C1$ to  $O2$, while transition rates determine other state transitions.\par

\subsection{Master Equation}
Under the principle of mass action kinetics, a differential equation called the master equation describes the chemical kinetics of the receptor \cite{swain1984handbook}, \cite{channelrhodopsin2018}. The master equation takes two parameters: the rate matrix $Q$, and  the probability vector of a receptor in state $i$ at time $t$.\\
The rate matrix $Q = [q_{ij}]$ represents a $k \times k$ matrix of {\em per capita} transition rates, where diagonal entries are set to make the sum of the row zero and $q_{ij}$ represents instantaneous transition rate from state $i$ to state $j$. If we visualize the rate matrix $Q$ using a graph, states are represented by $k$ and a directed edge is drawn from vertex $i$ to $j$ if $q_{ij} > 0$. In this case, we assume that rates that are affected by the input are directly proportional to $x(t)$ from \cite{channelrhodopsin2018}.  For example, $q_{12}x(t)$ is the transition rate from state $C1$ to state $O2$, which is sensitive, while $q_{23}$ is the transition rate from  $O2$ to $D3$, which is insensitive. Fig. \ref{fig2}a shows that for ChR2, where $k=3$, the transition rate $q_{12}= q_{12}x(t)$ as it varies with input $x(t)$ while rates $q_{23}$ and $q_{31}$ remain constant.
\rev{This is our key assumption to show the effect of the input signal on the system. There are systems in which signals act non-linearly on rates; however, these systems are not the focus of the discussion in this paper.}

For a receptor with $k$ discrete states, there exists a $k$-dimensional vector of state occupancy probabilities $p(t)$, given by \cite{channelrhodopsin2018}
\begin{align}
   \label{eqn:state probabilities}
     p(t) &= [p_1(t),p_2(t),. . ., p_k(t)] ,
 \end{align}
where $p_i(t)$ represents the probability of a given receptor occupying state $i$ at time $t$. The environmental conditions at the receptor, such as light level are known as the input $x(t)$.

Using $Q$ and $p(t)$ , the master equation is given by
\begin{align}
   \label{eqn:master equation}
     \frac{dp(t)}{dt} &= p(t)Q \,.  
 \end{align}

Table I shows the rate parameter values from the literature which we use in our model. \rev{The rates are measured in units of {\em transitions per second}. However, in the case of $q_{12}$, the units are {\em transitions per second per lumen}; since $x(t)$ is measured in {\em lumens}, the rate $q_{12}x(t)$ is again measured in transitions per second.}  \\

\begin{table}[ht!]
    \centering
    \begin{tabularx}{0.5\textwidth} { 
  | >{\raggedright\arraybackslash}X 
  | >{\centering\arraybackslash}X 
  | >{\raggedleft\arraybackslash}X | }
  \hline
  Parameter & Rate values & Units \\
  \hline
  $q_{12}$x(t) & $(5\times10^3)$x(t) & $s^{-1}$\\
  \hline 
  $q_{23}$     & $50 $       &      $s^{-1}$\\
  \hline
  $q_{31}$     & $17$        &        $s^{-1}$\\
  \hline
\end{tabularx}\\
    \caption{Rate Parameters for ChR2 from \cite{nagel2003channelrhodopsin}}
    
    \label{tab:Table I}
\end{table}

\rev{Based on the definition from \cite{channelrhodopsin2018}, the rate matrix $Q$ is expressed as\\\\
\begin{equation}
  \begin{bmatrix}
 -q_{12}x(t)& q_{12}x(t) & 0\\
   0 & -q_{23} & q_{23}\\
   q_{31} & 0 & -q_{31}
\end{bmatrix} 
\end{equation}\\
}
\rev{The diagonal elements in the matrix which represent self transitions are set to the negative of the corresponding transition rates. They make row sum equal to zero. Moreover, $q_{13}$, $q_{21}$ and $q_{32}$ are the rate of transition in reverse direction. Hence, our model does not support transition in the reverse direction, so these rates are set to $0$ in the matrix.}


 \begin{figure}[t!]
\centering
\includegraphics[width=\columnwidth]{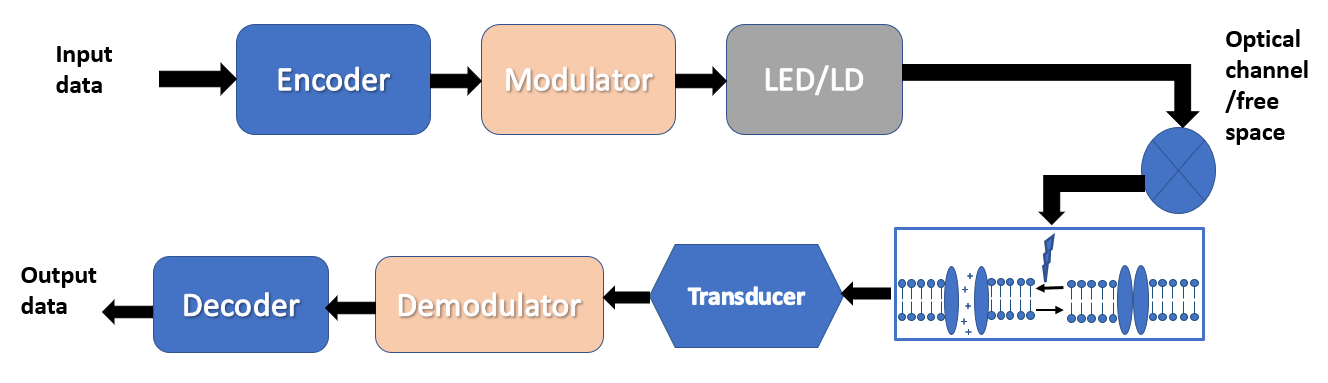}
\caption[\hspace{0.2cm}]{Block diagram of our proposed model. The transmitter consists of an encoder, modulator, and light source, whereas on the receiver side, a photoreceptor, channelrhodopsin, is used, followed by the transducer and decoder.
Figure adapted from \cite{idris2019visible}.}
\label{fig1}
\end{figure}

\begin{figure}[h!]
\centering
\includegraphics[width=\columnwidth]{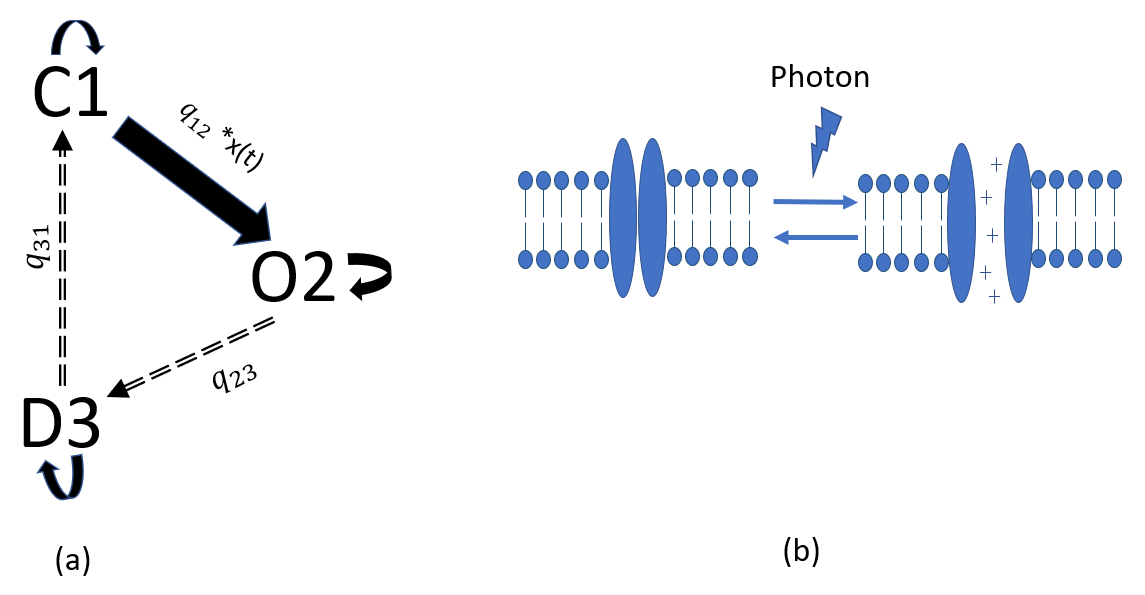}
\caption[\hspace{0.2cm}]{Three state model: (a) closed/ground state $C1$, open state $O2$ and desensitized/intermediate state $D3$. $q_{12}x(t)$, $q_{23}$ and $q_{31}$ are transition rates from transition matrix \ref{eqn:transition matrix} where x(t) is light intensity (b) ChR2 symbol showing that light changes its state from closed to open. Figure adapted from \cite{nikolic2009photocycles}.}
\label{fig2}
\end{figure}

\subsection{Discrete-time  model}
Continuous-time communication channel analysis using the continuous-time master equation is mathematically challenging. As a result, we propose the discrete-time, discrete-state channel, which may be rigorously analyzed, to study its properties with a fixed time interval $\Delta t$. 
Signal transduction receptors such as channelrhodopsin can be seen as finite-state, intensity-modulated  Markov chains in which the transition rates between certain pairs of states are sensitive to the input (though other transitions may be independent of the input) \cite{channelrhodopsin2018}.

The master equation can be approximated in discrete time as 
\begin{align}
   \label{eqn:discrete time}
     \frac{dp}{dt} &= \frac{(p(t+\Delta t))-p(t)}{\Delta t} + \rev{o(\Delta t)} = p(t)Q,\  as\  \Delta t\rightarrow 0\,, \\
      p(t+\Delta t) &= \Delta t p(t)Q(t) +p(t) + o(\Delta t)\,,\\
                      &= \Delta t p(t)Q(t) +p(t)I + o(\Delta t)\,,\\
                      &= p(t)(I + \Delta t Q)+o(\Delta t)\   \text{as}\   \Delta t\rightarrow 0\,,
\end{align}
\rev{where $o(\Delta t)$ represents a term that is negligible as $\Delta t \rightarrow 0$.}
 For a discrete-time model, we introduce the approximation
 \begin{align}
   \label{eqn:discrete time2}
     p_i = p(i\Delta t)+o(\Delta t),\ as\ \Delta t \rightarrow 0\,,\\
     p_{i+1}=p_i(I+\Delta t Q)\,,
\end{align}


Based on \cite{channelrhodopsin2018}, we describe the probability transition matrix for discrete-time Markov chains using  as follows:
 \begin{align}
   \label{eqn:transition matrix}
   P &= I+\Delta t Q\,.
 \end{align}
 We will briefly explain the variables used in the equation:
 \begin{itemize}
     \item $Q$ is the rate matrix defined in the previous section
     \item $I$ is the identity matrix of the appropriate size. 
     \item $\Delta t$ is a  time step introduced to study the properties of the receptor as a discrete-time, discrete-state channel influenced by a continuous time channel. The time step $\Delta t$ should be small enough so that $P$ meets the conditions of the Markov chain transition matrix (row-stochastic and nonnegative).
 \end{itemize}
   The state transition matrix for ChR2 is \\\\\\
\begin{equation}
 \label{eqn:transition matrix values}
  P = \begin{bmatrix}
1 & 0 & 0\\
0 & 1 & 0\\
0 & 0 & 1
\end{bmatrix}
+\Delta t
\begin{bmatrix}
 -q_{12}x(t)& q_{12}x(t) & 0\\
0 & -q_{23} & q_{23}\\
q_{31} & 0 & -q_{31}
\end{bmatrix}
\end{equation}\\

%

\section{Communication model}
Our goal is to demonstrate a novel communication system using  the biological receptor ChR2. We will first analyze using a single receptor, then extend to multiple receptors. In the following, we describe how to view ChR2 as the receiver in a light-based communication system.

\subsection{Communication system components}
The components of the communication model in Fig. \ref{fig1} are described as follows.
\begin{itemize}

    \item {\em Transmitter.} The transmitter consists of an encoder, modulator and light source. The input bit stream is first passed through a line encoder to yield an encoded bit stream, then modulation is performed, and data is fed to the light source for transmission through the optical channel. We follow the setup in \cite{bamann2008spectral} and assume that a laser source (400 nm) is used to excite the ChR2.
     
    \item {\em Channel.}
    The channel is an environment through which light can propagate. It can be free space or any other optical channel, such as optical fibre.
     \item {\em Receiver.}
     Generally, a photodiode is used for the reception of VLC signals \cite{khan2017visible}. However, we have replaced it with the biological receptor (ChR2), in which a photocurrent is produced when blue light is illuminated. We can use any current measuring instrument to detect current. By using a demodulator and decoder, the transmitted message can be detected.

\end{itemize}

    
\subsection{Stochastic channel model}
We will now discuss the channel input and output state relationship of receptors. The matrix $P$ is a non-negative and row-stochastic matrix that meets the conditions of a Markov chain transition probability matrix as long as observation time $\Delta t$ is small enough (\ref{eqn:transition matrix values}). We will use the same notations as in \cite{channelrhodopsin2018}.\\
\begin{itemize}
    \item {\em Input:} As mentioned in Section II, the transition of the receptor from state $C1$ to state $O2$ is sensitive to the intensity of light $x(t)$. Input bits are transmitted using the low and high values of $x(t)$ with different values of time step $\Delta t$. However, values of Table \ref{tab:Table I} are determined for $\Delta t = 1\times 10^{-6}$. We use on-off keying modulation, described below.
    \item {\em States}: The output in this case is the state of the receptor represented by $s_{i}$. As in Fig. \ref{fig2}a, the receptor may be in one of three states, the channel state $s_{i}$ is represented by a number and is also discretized. Hence, the output vector would be $\textbf{s}=[s_{1},s_{2},...,s_{n}]$.
    \item {\em Input-State response}: The relationship between channel input and output can be described as a PMF
     \begin{align}
        \label{eqn:psgivenx}
         p(\textbf{s}|{x}) & = \prod_{i=1}^{n}p(s_{i}|x_{i},s_{i-1})\,.
     \end{align}
    where $p(s_{i}|x_{i},s_{i-1})$ is given by transition matrix P. The transition from state $s_{i-1}$ to $s_{i} $ is considered \emph{insensitive} if for all $x_{i}\in \chi$, $p(s_{i}|x_{i},s_{i-1})= p(s_{i}|s_{i-1})$. In our case, if the transition from state $C \rightarrow O$ is sensitive to input then the \emph{self transition} $C\rightarrow C$ is also sensitive to input. 
\end{itemize}

\subsection{Modulation technique}

    
 The state diagram of ChR2 shows three states, but the response of the desensitized state and the closed state are electrically equivalent, i.e., neither allows a photocurrent to pass. When the channelrhodopsin is illuminated, the state can change from $C1$ to $O2$ ($C1 \rightarrow O2$), and ion current is detected. The transition from state $O2 \rightarrow D3$ depends upon transition rate $q_{23}$ mentioned in rate matrix $Q$. We will use these properties to design our modulation technique for a communication system.
 
 For simplicity, to select the channel input $x(t)$, we consider on-off keying modulation. The transmission may be represented by light pulses remaining {\emph{ on} or \emph{off}} for a specific amount of time to transmit a sequence of bits $\vec{b} = [b_1,b_2,\ldots]$, where $b_i \in \{0,1\}$ is the bit transmitted in the $i$th interval. Hence, $(x(t)=0)$ represents input bit '0', and the input bit '1' is represented by ($x(t)=1$) (in units corresponding to the rate in Table \ref{tab:Table I}).
    
Equivalently, it will be convenient to represent  $\vec{b}$ as a vector of pulses. For example, a light of frequency $ ~440$ nm is generated for $10$ ns seconds to send bit 1. The detector observes the receptor and records its conditions in $n$ regular intervals of $\Delta t$. If there is a current or a state transition between $C1$ and $O2$, it is interpreted as bit 1; otherwise, as bit 0. $n$ is the number of times we observe the output of the receiver before making a decision whereas $ T= n\Delta t$ is total time.
Suppose the transmitter releases $m$ bits. The input pulses for these bits can be represented by $\Vec{x}=[x_1(t),x_2(t),...x_m(t)]$; for example, if $b_i = 1$ then the release time of $x_i$ which is a light pulse is $iT$.


\begin{figure}[ht!]
\centering
\includegraphics[width=8.5cm]{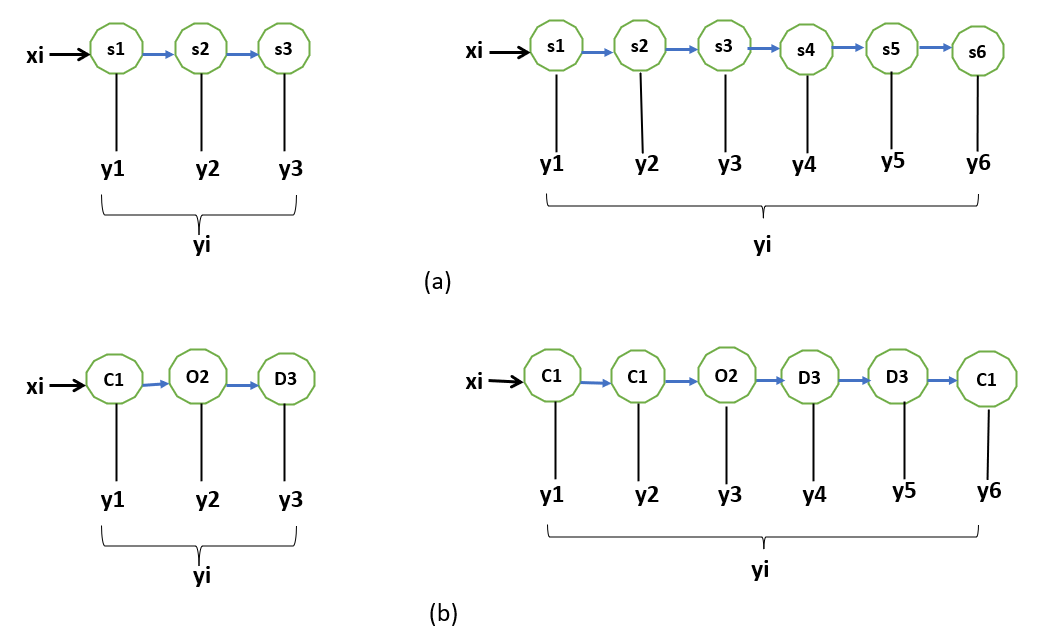}

\caption{State combination example:(a) Input $x_{i}(t)$ causes state changes which is measured in interval of time $\Delta t$, output $yi$ is detected after observing each output after sampled interval,(b) is an example of state combinations of ChR2}
\label{Fig4}
\end{figure}
\subsection{Receiver and Detection}
 In our proposed receiver, ChR2 remains {\em closed} without light and becomes {\em open} as light illuminates it after a random amount of time. In terms of detection, we attempt to detect bit $i$ as a vector $\Vec{y_i}$, thus the vector ${\Vec{y}_i}={y_1,y_2,y_3,...y_n}$ is an output of each state observed within a discrete-time interval $\Delta t$. The decision is based on the  aposterori probabilities calculated for each state combination given in Table II for $n=3$. Fig. \ref{Fig4}a and b illustrate the detection mechanism for the three and six-state combinations. Three observations represent the minimum requirement due to the presence of three states, whereas the choice of $n=6$ is made for comparison purposes. As an example, consider $n=3$, i.e., the detector will observe the receptor for $3\Delta t$ times. There are $3^3=27$ combinations of states, among which only $12$ are possible based on the presence of the current. The decision rule is determined based on {\emph{a posteriori probability}} calculated for each state combination:
\begin{align}
 \label{eqn:aposteriori}
     p(x|\textbf{y}) &= \frac{p(\textbf{y}|{x})p(\textbf{x})}{\sum_x p(\textbf{y}|\textbf{x})p(\textbf{x}) }\,,\\
                              &= \frac{\sum_s\prod_{i=1}^{n} p(y_i|s_i)p(s_i|s_{i-1},x)p(x)}{\sum_x \sum_s\prod_{i=1}^{n}p(y_i|s_i)p(s_i|s_{i-1},x)p(x)}\,, 
 \end{align}
 where
\begin{align}
     \label{eqn:pygivens}
     p(y|s) &= 
     \left\{
        \begin{array}{cl}
             1, & y=1,\,s=O\,; \\
             1, & y=0,\,s=C\, \text{or}\, D\,; \\
             0, & \text{otherwise}
        \end{array}
    \right.
\end{align}
and $p(s|x)$ can be determined from (\ref{eqn:psgivenx}) using transition matrix $P$ from (\ref{eqn:transition matrix}). Finally, the decision rule is described as (with the notation $a \rightarrow b$ signifying ``if $a$ occurs, then decide $b$''):
 \begin{align}
     \label{eqn:comparison1}
     p({x=0}|\textbf{y}) > p({x=1}|\textbf{y})\rightarrow 0\,,\\
     \label{eqn:comparison2}
     p({x=0}|\textbf{y}) < p({x=1}|\textbf{y})\rightarrow 1\,,\\
     \label{eqn:comparison3}
     p({x=0}|\textbf{y}) = p({x=1}|\textbf{y})\rightarrow \epsilon\,.
\end{align}

Table II shows observable state combinations and their probabilities calculated from (\ref{eqn:aposteriori})-(\ref{eqn:pygivens}).
As an example of our detection mechanism, suppose the detector observes {\em no current} in the first interval $\Delta t$ and again observes {\em no current} in the second interval; however, in the third interval, it detects {\em current} so the $\vec{y} = [0,0,1]$ as an output $y_1,y_2,y_3$. Table II shows that this output corresponds to (D3, C1, O2) or (C1, C1, O2). The transition rates from Table \ref{tab:Table I} are used to derive the probabilities. Based on the decision rule, (\ref{eqn:comparison1}-\ref{eqn:comparison3}) the detector will decide in favour of bit '1' assuming that by turning on the light, the transmitter has sent bit '1'.\par

Consider another example of sending bit '0' and the detector observes {\em an ion current} in the first interval and {\em no current} in the other two intervals so that the corresponding output will be considered as $\vec{y}=[1,0,0]$. According to rows (11-14) in Table II, $p(x=0|y) = p(x=1|y)$, so the message will be undetermined, and the receiver will display an error message. On the other hand, if the detector observes {\em no current} for three discrete time intervals, then it indicates (C1, C1, C1), (D3, D3, D3), (D3, D3, C1), (D3, C1, C1) that are $\vec[y]=0,0,0$, so the message will be considered as bit '0' as $p(x=0|y) > p(x=1|y)$. Some combinations like 
(O2, C1, D3), (D3, O2, C1)  is practically impossible, so their probabilities cannot be determined. The probabilities can be determined for any number of states, and here we have considered sequences for three and six states which are observed in $\Delta t$ steps for the total time interval of $T$.

\begin{table}[!h]
    \centering
    \begin{tabularx}{0.5\textwidth} { 
  | >{\raggedright\arraybackslash}X 
  | >{\centering\arraybackslash}X 
  | >{\raggedleft\arraybackslash}X | }
  \hline
  State Combinations & $p(x|y),x=1$ & $p(x|y),x=0$ \\
  \hline
  C1-C1-C1 & $0.34$ & $0.66$\\
  \hline 
  D3-D3-D3    & $0.34 $       &      $0.66$\\
  \hline
  D3-D3-C1    & $0.34 $       &      $0.66$\\
  \hline
  D3-C1-C1    & $0.34 $       &      $0.66$\\
  \hline
  D3-C1-O2     & $1.0$        &        $0$\\
  \hline
  C1-C1-O2     & $1.0$        &        $0$\\
  \hline
  C1-O2-O2     & $1.0$        &        $0$ \\
  \hline
  C1-O2-D3     & $1.0$        &        $0$\\
  \hline
  O2-O2-O2     & $0.5$        &        $0.5$\\
  \hline
  O2-O2-D3     & $0.5$        &        $0.5$\\
  \hline
  O2-D3-D3     & $0.5$        &        $0.5$\\
  \hline
  O2-D3-C1     & $0.5$        &        $0.5$\\
  \hline
  other combinations & impossible  & impossible\\
  \hline
\end{tabularx}
    \caption{A posteriori probabilities of states when $n=3$. The transition rates $q_{12}x(t), q_{23}$ and $q_{31}$ are used to determine these probabilities}
    
    \label{tab:Table II}
\end{table}

 Data rates can be determined by
\begin{align}
    \label{eqn:DataRate}
    R = \frac{1}{T}\,. 
\end{align}
in bits per second. This simplified rate is sufficient to demonstrate the feasibility of the scheme. We leave the more sophisticated design to future work.

\subsection{Multiple Receptors}
So far, we have defined our model by assuming only one channelrhodopsin receptor (ChR2). In this section, we will extend our analysis to multiple receptors, which are \rev{ {\em independent, statistically independent and indistinct}.} 
 We take an example of two receptors for illustration. The purpose of taking multiple receptors into account is to reduce the waiting time, hence increasing the data rate since there is more likelihood of opening of at least one receptor. In order to obtain the {\em transition probability matrix}, we have determined possible combinations of two receptors in three states, as shown in Table III. The first row of the table contains the number of states $C1$, $O2$, and $D3$, whereas the other rows indicate the number of receptors in each state. Each combination of states is represented by alphabetical letters in the last column of Table III.\par The state table determines transitions of receptors from one state to another, leading to the transition probability matrix. To find the total number of possible combinations, the {\em star and bar method} from \cite{feller1991introduction} can be used. Using this method, if there are ${N}$ items and $k$ bins, then $k-1$ bars or separators are required to get $N$ items into $k$ bins. There are $(N+k-1)$! permutations of ordering $n$ items. The permutation of $n$ items does not matter and the permutation of $(k-1)$ separators does not matter.
\begin{figure}[ht!]
\centering
\includegraphics[width=\columnwidth]{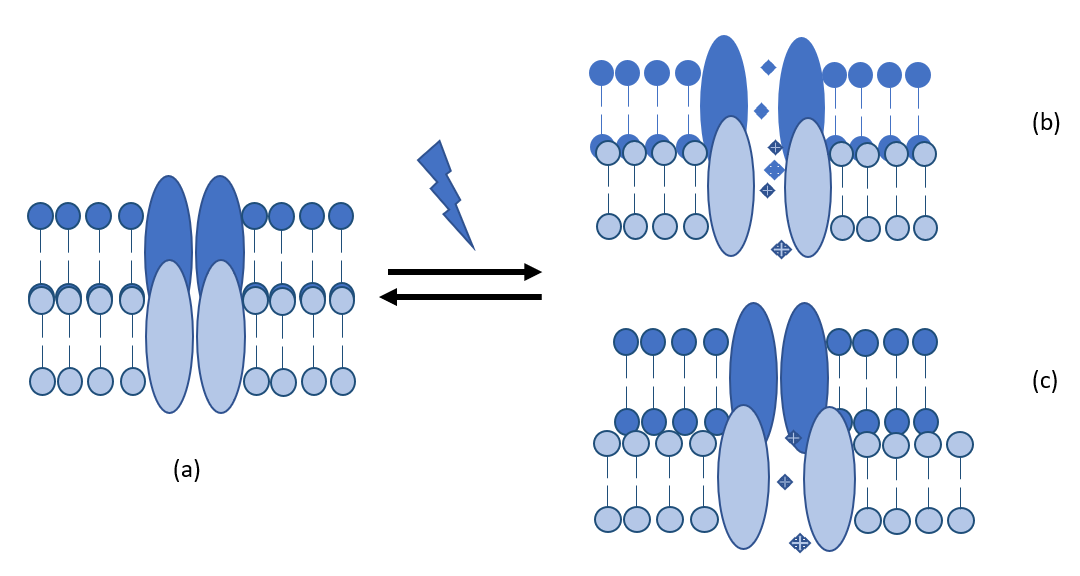}
\caption{ Three ChR2 receptors for illustration. (a) All receptors are in closed states (b) After illumination of blue light, two receptors are open (c). Only one receptor is open.}
\label{fig2ChR2}
\end{figure}
 Thus we have
                     
\begin{align}
    \label{eqn:combination}
    \binom{N+k-1}{k-1} = \frac{(N+k-1)!}{N!(k-1)!}\,.
\end{align}
Based on (\ref{eqn:combination}), if we have $N=2$ ChR2 receptors and $k=3$ states, then there are six combinations, as listed in Table III. Following the process, the state transition diagram can be determined by considering the possibilities of transitions from one state combination to another.\\

\begin{table}[ht!]
    \centering
    \begin{tabularx}{0.5\textwidth} { 
  | >{\raggedright\arraybackslash}X 
  | >{\centering\arraybackslash}X 
  | >{\raggedleft\arraybackslash}X 
  | >{\raggedleft\arraybackslash}X| }
  \hline
  C1 & O2  & D3 & combination state \\
  \hline
  $2$ & $0$ & $0$ & A\\
  \hline 
  $1$  & $1$  &  $0$ & B \\
  \hline
  $1$   & $0$  &  $1$ & C\\
  \hline
  $0$  & $2$   &  $0$  & D\\
  \hline
  $0$   & $1$  &  $1$ & E\\
  \hline
  $0$  & $0$   &  $2$  & F\\
  \hline
\end{tabularx}\\
    \caption{ combinations of two receptors in three states}
    \label{tab:Table III}
\end{table}

 As an example, if we have two receptors in combination state A in Table III  then there is a chance that one of the receptors changes its state from $C1$ to $O2$ that is combination state B in Table III.
 \begin{figure}[h!]
\centering
\includegraphics[width=0.9\columnwidth]{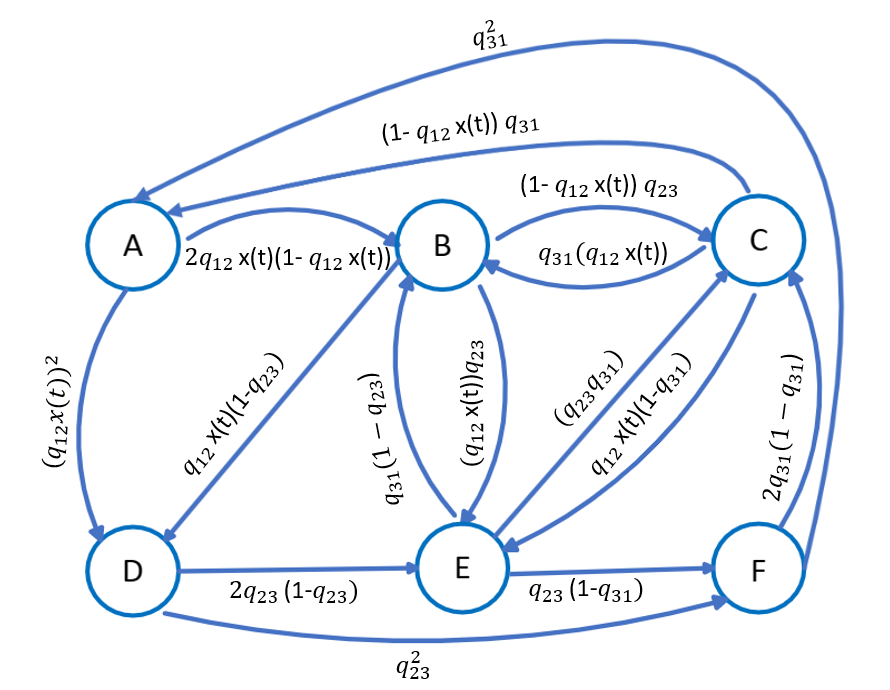}
\caption[\hspace{0.2cm}]{Directed graph showing possible states and their transition probabilities between states when there are two receptors.}
\label{figtransition}
\end{figure}
 In the same way, both receptors can change their states to $O2$. However, they cannot go to Desensitized state $D3$ directly from $C1$ so all other combinations are not possible. The state diagram  is shown in Fig. \ref{figtransition}
The transition probabilities are determined by using rate matrix Table \ref{tab:Table I}. For instance, the transition probability from combination A to D will be $(q_{12}x(t))^2$ because both receptors can move to that state where $q_{12}x(t)$ is the probability of transition from state $C1$ to $O2$

The probability transition $P_c$ matrix can be determined from the state diagram showing all possible transitions from one combination state to another. 



\begin{figure*}[!h]
\normalsize

\setcounter{equation}{20}
\begin{equation}
\label{eqn_dbl_y}
 P_c=
  \begin{bmatrix}
(q_{12}x(t)-1)^2&{\alpha}&0&{(q_{12}x(t))^2}&0&0\\
  0 & {\beta}&{(1-q_{12}x(t))q_{23}}&{(q_{12}x(t))(1-q_{23})}& {(q_{12}x(t))q_{23}}&0\\
  {(1-q_{12}x(t))q_{31}}& q_{31}q_{12}x(t)& {\gamma}&0& q_{12}x(t)(1-q_{31})&0 \\
  0&0&0&(q_{23}-1)^2&{2q_{23}(1- q_{23})}&(q_23)^2 \\
  0& q_{31}(1-q_{23})& q_{23}q_{31}&0& 1-{q_{31}(1-q_{23)}+q_{23}}& q_{23}(1-q_{31}) \\
  q_{31}^2 & 0&2q_{31}(1-q_{31}) &0 &0&(q_{31}-1)^2\\
\end{bmatrix}
\end{equation}
\hrulefill
\vspace*{4pt}
\end{figure*}

Some of the entries in matrix $P_c$ are represented by $\alpha$, $\beta$ and $\gamma$ to optimize the space, where
\begin{align}
    \label{eqn:substitute}
    \alpha &= {2(1-q_{12}x(t))q_{12}x(t)}\,,\\
    \beta &= 1-{(1-q_{12}x(t))q_{31}+q_{12}x(t)}\,,\\
    \gamma &= 1-{(1-q_{12}x(t))q_{31}+q_{12}x(t)}\,.
\end{align}
 However, they can be easily determined from the state diagram in Fig. \ref{figtransition}.\par
The generalized process of generating a probability transition matrix for any number of multiple receptors can be described as follows.
\begin{itemize}
    \item For $N$ channelrhodopsin receptors, each with three discrete states $C1$, $O2$ and $D3$, calculate the number of combinations using (\ref{eqn:combination})
    \item Make a state table with four columns: three columns with the names of individual states and a fourth column with the names of state combinations. Fill in each row with a receptor as illustrated in Table III, showing different numbers of receptors in different states 
    \item Draw a state diagram by connecting all possible state combinations in the form of a directed graph where each edge corresponds to the edge of the rows and arrows indicate transition probabilities.
    \item Create a probability transition matrix based on the rate matrix with rows and columns equal to the number of state combinations calculated from (\ref{eqn:combination}).
    
\end{itemize}
\subsection{Noise Model}
In Section III A, we considered a laser source that emits a perfectly coherent beam of light with high intensity to activate the ChR2 channel with the probability $1$ if the present state of the receptor is $C1$. 

However, using high-intensity light can saturate the ChR2 molecules. On the other hand, low-intensity light is more vulnerable to photon fluctuations. There are statistical fluctuations on short-time scales due to the discrete nature of the photons. As light intensity is proportional to the number of photons in a given interval per unit area, these fluctuations cause light intensity to deviate from its constant value.
\begin{align}
\label{eqn:intensityandphoton}
    x(t) &= \frac{f}{A}E\,,\\
         &= fX\,,
\end{align}
Here, $f$ is the number of photons in a given interval of time, area $A$ and energy $E$ are constant so their ratio has been replaced by $X$.
 The probability of finding the $f$ number of photons in a given interval of time is determined by Poissonian photon statistics \cite{photonstatistics} as
\begin{align}
\label{eqn:poisson}
    P(f) &= \frac{\lambda^f}{f!}\exp{(-\lambda)}\,,
\end{align}
where $\lambda$ is the average number of photons during an interval of time.\par 
Signal to noise ratio (SNR) can be calculated as
\begin{align}
\label{eqn:SNR}
    SNR &= \frac{\text{Signal power}}{\text{Noise Power}}\,,
\end{align}
Poisson distributions are uniquely characterized by their mean value $\lambda$. The fluctuations of a statistical distribution about
its mean value are usually quantified in terms of the variance. The variance is equal to the square
of the standard deviation $\Delta f$ and is defined by
\begin{align}
    (\Delta{f})^2 &= \sum_{f=0}^{\infty}(f-\lambda)^2P(f)\,,
\end{align}
 For Poisson statistics the variance is equal to the mean value:
\begin{align}
    (\Delta f)^2=\lambda\,,
\end{align}

The standard deviation for the fluctuations of the photon number above and below the mean
value is therefore given by
\begin{align}
 \Delta{f}=\sqrt{\lambda}\,,   
\end{align}
This shows that the relative size of the fluctuations decreases as $\lambda$ gets larger. So, the noise power is proportional to the fluctuation, which is the square root of the average photon count, whereas
the optical signal power is proportional to the average number of photons $\lambda$. Hence
\begin{align}
\label{eqn:optical noise}
    SNR &= \frac{\lambda}{\sqrt{\lambda}}\,,\\
        &= \sqrt{\lambda}\,.
\end{align}

\section{Results and Discussion}
A computer simulation was conducted to evaluate the performance of our proposed system based on Fig. \ref{fig1}. We use full Monte Carlo simulations to obtain the probability of error and data rates for different input intensities and $\Delta t$ values. \rev{The values of parameters used in our simulations are listed in Table IV}. We have also calculated the probability of error to verify it with respect to $\Delta t$ The probability of error versus total time is illustrated with semilog plots to visualize smaller values better.

\begin{table}[h!]
\centering
\begin{tabular}{  m{14em}  } 
 \hline
  Parameters used in the simulation of communication systems  \\
  \hline
  $n = 3$\\
  $n = 6$\\
  $\Delta t =$ $3$ms, $7$ms, $10$ms\\\\
  
  number of ChR2 = $1$, $3$, $5$, $7$\\
  \hline 
  \end{tabular}\\
\caption{Parameters of communication systems}
    \label{tab:Table IV}
\end{table}
\rev{\subsection{Theoretical verification of the probability of error}
In order to verify the simulation result, we have calculated the probability of error as 
\begin{align}
\label{eqn: probability of error calc}
    P_e = p_e(y=1|x=0)p(x=0)+p_e(y=0|x=1)p(x=1),\\
    p(y=1|x=0)= \sum_s\prod_{i=1}^{n} p(y=1|s_i)p(s_i|s_{i-1},x=0),\\
    p(y=0|x=1)= \sum_s\prod_{i=1}^{n} p(y=0|s_i)p(s_i|s_{i-1},x=1).
\end{align}
Here, $P_e$ is the probability of error, which is calculated using the parameters in Table II.
} 

\begin{figure}[h!]
\centering
\includegraphics[width=\columnwidth]{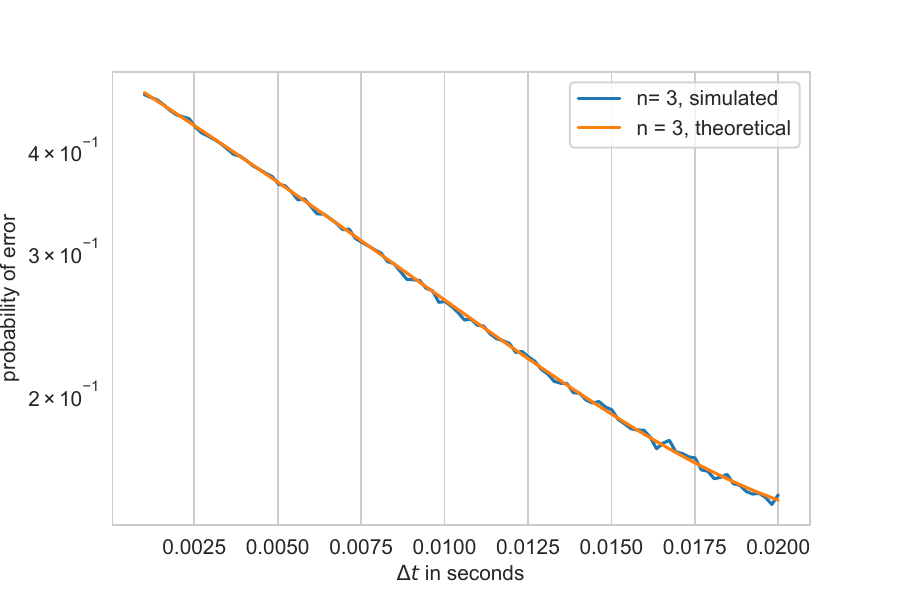}
\caption{Probability of error vs $\Delta t$
$n=3$. The red line shows the theoretical probability of error, while the blue line is the empirical formulation of the error probability  }
\label{figpb}
\end{figure}
 \rev{As mentioned in Section III, the total time $T$  is the interval in which the detector keeps observing the state of ChR2 receptors with a time step of $\Delta t$ whereas $n$ is the length of the sequence of observations. We will use $n=3$ as our simplest case to verify the probability of error. However, in our subsequent sections, we will express the effects of the number of observations on data rates.}\par
 \rev{The probabilities of error are the sum of two conditional probabilities defined as the probability of receiving incorrect information. For example, $p(y=1|x=0)$ shows the chance of receiving bit $1$ when bit $0$ is transmitted. Our biomolecular receiver is modelled as a finite-state Markov chain where output $y$ is dependent on states $s_i$ and states $s_i$ are dependent on input $x_i$. Equations (\ref{eqn: probability of error calc}) show the relationship between $y$, states $s_i$ and input $x$ in terms of conditional probabilities. As an example, consider the case of (35), when $n=3$, the sequence of observing states is $s_1,s_2,s_3$, which could be any state listed in the first column of Table II. In order to determine $p(s_1,s_2,s_3|x)$, we have calculated {\em trajectory probabilities} using transition matrix (\ref{eqn:transition matrix values}) starting with initial states selected on the basis of the steady-state distribution.} \par

 \rev{Fig. \ref{figpb} shows that the probability of error found theoretically using (\ref{eqn: probability of error calc}) is perfectly aligned with the probability of error obtained from simulation. For the sake of simplicity, rigorous mathematical calculations will not be provided for each subsequent result in the forthcoming sections. Nevertheless, the results from Fig. \ref{figpb} allow us to conclude that our simulation results are accurate. }
 
\subsection{Single Receptor}
We present our results in terms of the probability of error as a function of total time $T$. As mentioned in Section III, total time $T$  is the interval in which the detector keeps observing the state of ChR2 receptors with a time step of $\Delta t$ whereas $n$ is a number of observations that could be $3,6,...$. We will explicitly express the effects of the number of states on data rates. We have designed our system  for observing the receptor up to $3\Delta t$ ($n=3$) and $6\Delta t$ ($n=6$) time intervals  with different values of $\Delta t$.

\begin{figure}[ht!]
\centering
\includegraphics[width=\columnwidth]{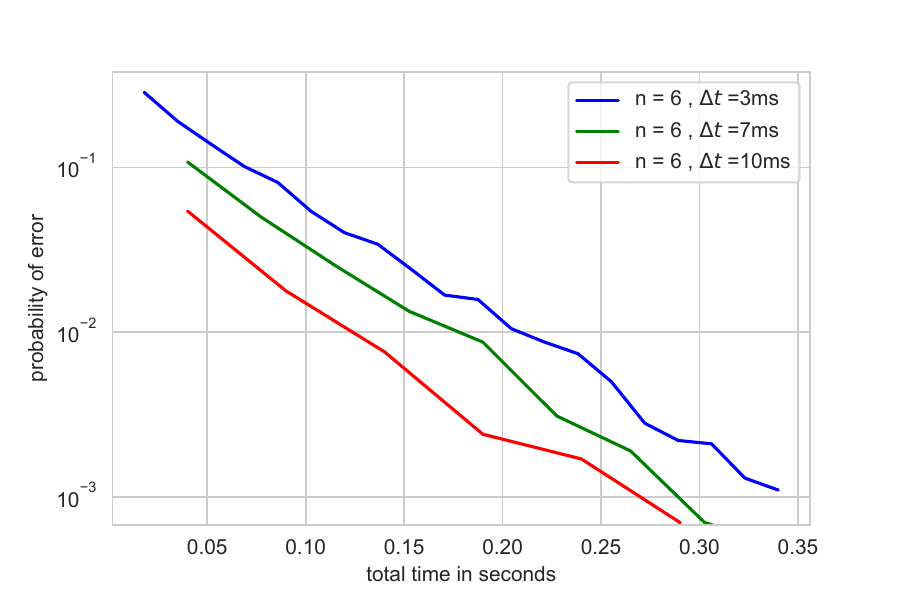}
\caption{Probability of error vs total time $T$ for $n=6$ at different $\Delta t$. }
\label{fig4}
\end{figure}

\begin{figure}[hbt!]
\centering
\includegraphics[width=\columnwidth]{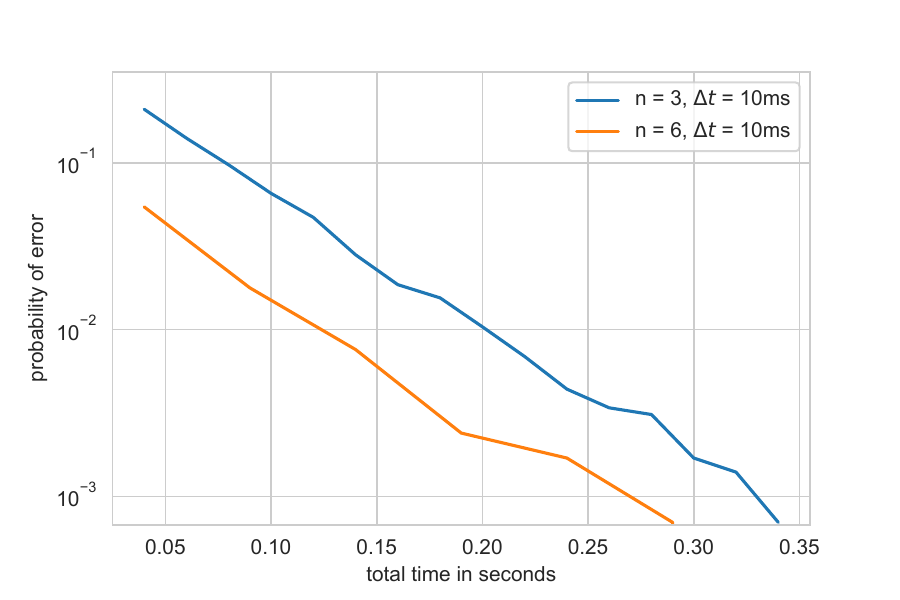}
\caption{Probability of error decreases as total time increases. }
\label{fig5}
\end{figure}

 Fig. \ref{fig4} depicts the effect of total time on the probability of error for state $n=6$ and for time steps $\Delta t=3,7,10$ms. It clearly shows that the probability of error decreases as the total time interval $T$ increases.
 Intuitively, When we spend more time observing there is a higher chance of getting the correct value of the state. Observation states, or how often the detector observes the receptor before making a decision, also affect the error rate. Fig. \ref{fig5} compares the error probability for states $n=6$ and $n=3$. Though it decreases the probability of error for a small total time $T$, it will increase the complexity of the system.


 In Fig. \ref{fig5}, we compare the probability of error for different observing states $n=3$ and $n=6$ with the time steps $\Delta t = 3$ms. The results show that the probability of error for states $n=3$ is initially higher $(12\%)$ than that of $n=6$.\par


\subsection{Data Rates} 





\begin{figure}[ht!]
\centering
\includegraphics[width=\columnwidth]{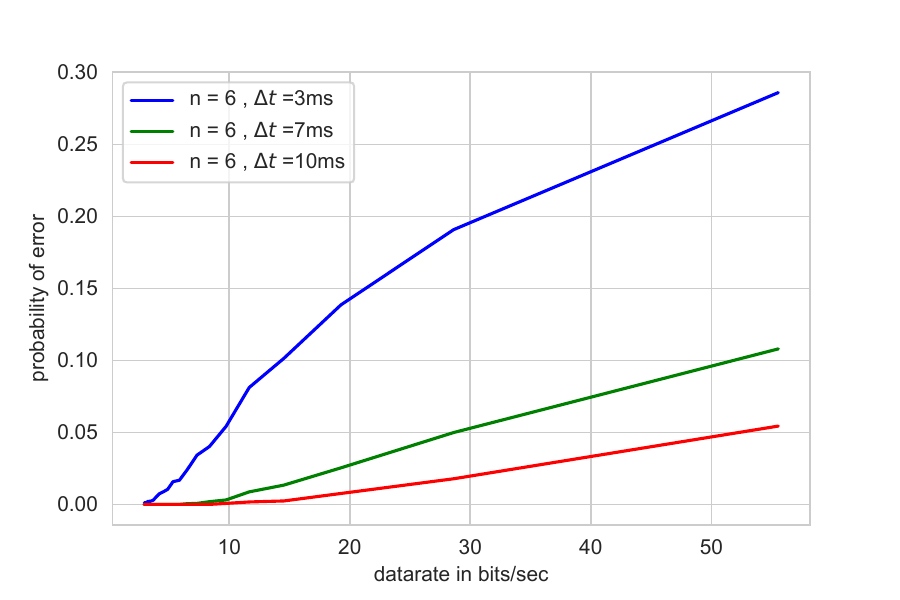}
\caption{Probability of error vs datarate for $n=6$ at different $\Delta t$. }
\label{fig8}
\end{figure}

Recall our simplified definition of data rate from (\ref{eqn:DataRate}). We have analyzed that there is a trade-off between data rates and the probability of error. Fig. \ref{fig8} shows estimated data rates with respect to the probability of error for $n=6$. Though the system has a low bit error rate, it experiences a low probability of error with an increase of $\Delta t$. As one example, in Fig. \ref{fig8} for $n=6$ the probability of error is more than $14\%$ to achieve the rate of $20$ bits/sec when $\Delta t = 3$ms. Similarly, it can be seen  that the probability of error increases with the data rate for the same $\Delta t$.

\subsection{Multiple Receptors: Improvement of Data Rates}
\begin{figure}[ht!]
\centering
\includegraphics[width=\columnwidth]{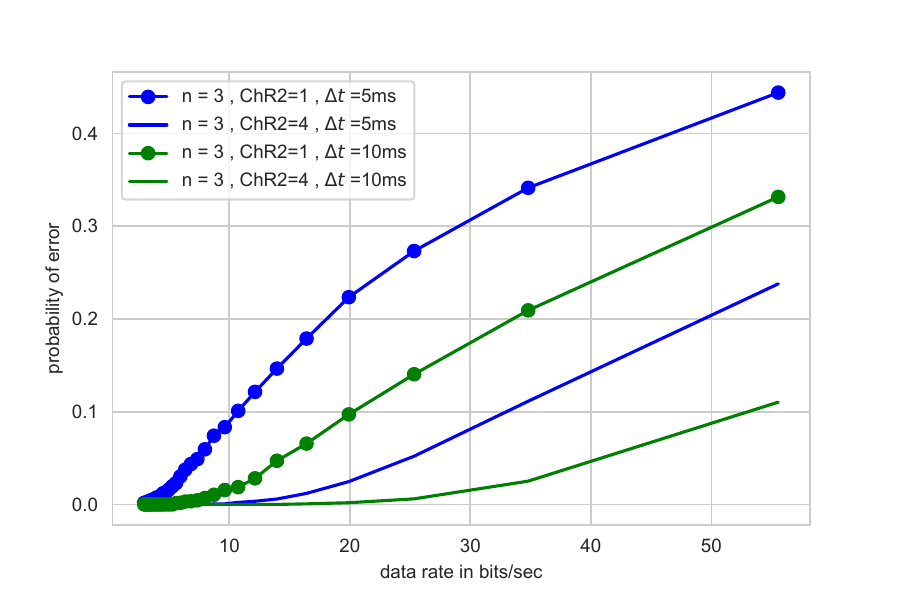}
\caption{Probability of error vs data rate for $1$ ChR2 receptor (marked line) and $4$ ChR2 receptors (lines without markers). The curves are plotted for different time steps and for $n=3$ time intervals. There is a significant decrease in probability of error for same data rates.}
\label{fig43ChR2}
\end{figure}

\begin{figure}[h!]
\centering
\includegraphics[width=\columnwidth]{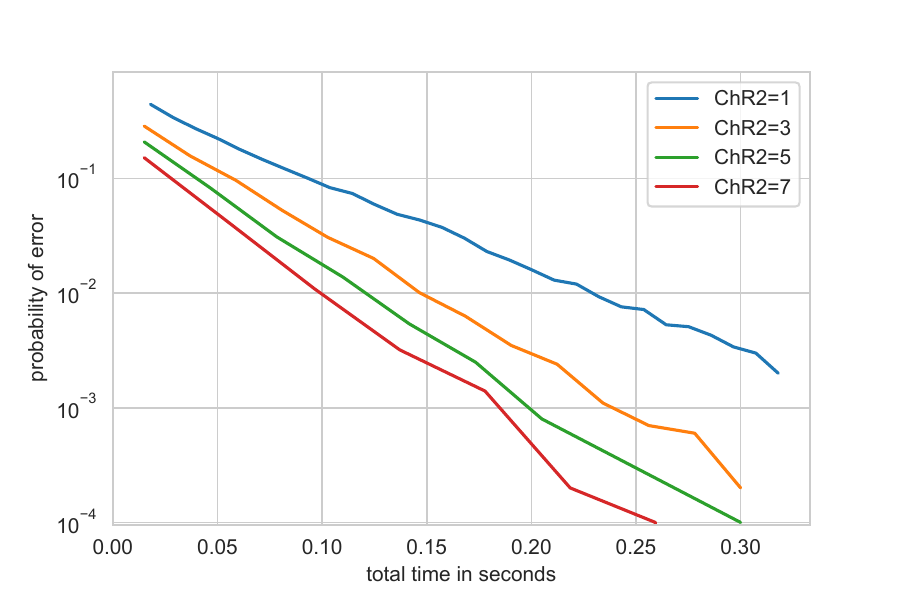}
\caption{Comparison of multiple ChR2 :Probability of error vs total time with time step of $5$ms. The probability of error is significantly reduced as the number of ChR2 increases. }
\label{figcomptandp}
\end{figure}

In this section, we will show that it is possible to obtain higher data rates by using multiple channelrhodopsin receptors instead of one. This can be seen in Fig. \ref{fig43ChR2} that compares the data rate of one channelrhodopsin receptor with that of four. The figure shows four curves; a line with a solid marker indicates error probability for one ChR2, whereas the smooth curves represent the probability of error vs the data rate for four ChR2. Each set of curves is drawn for the same parameter, with the same number of observing states $n=3$, and time steps $5$ ms and $10$ ms. It can be seen that the error probability is significantly lower for four ChR2 relative to one. This is further illustrated in Fig. \ref{figcomptandp}, which shows a comparison of one receptor with the receptors $3,5$ and $7$ in terms of error probability and total time.   We have used the same parameters as used in the case of a single receptor, that is, the time step of $5$ ms and state $n=3$. The figure shows that there is a significant decrease in the probability of error as compared to a single receptor. 


\begin{figure}[h!]
\centering
\includegraphics[width=\columnwidth]{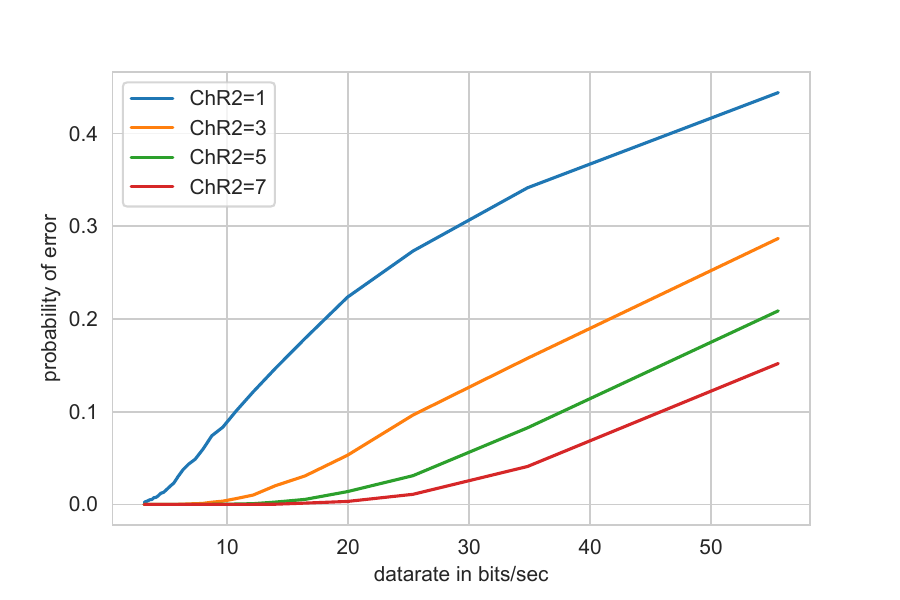}
\caption{Comparison of multiple ChR2 :Probability of error vs datarate with time step of $5$ms. As the number of ChR2 increases higher data rates can be achieved with a low probability of error. }
\label{figcompdandp}
\end{figure}

\begin{figure}[h!]
\centering
\includegraphics[width=\columnwidth]{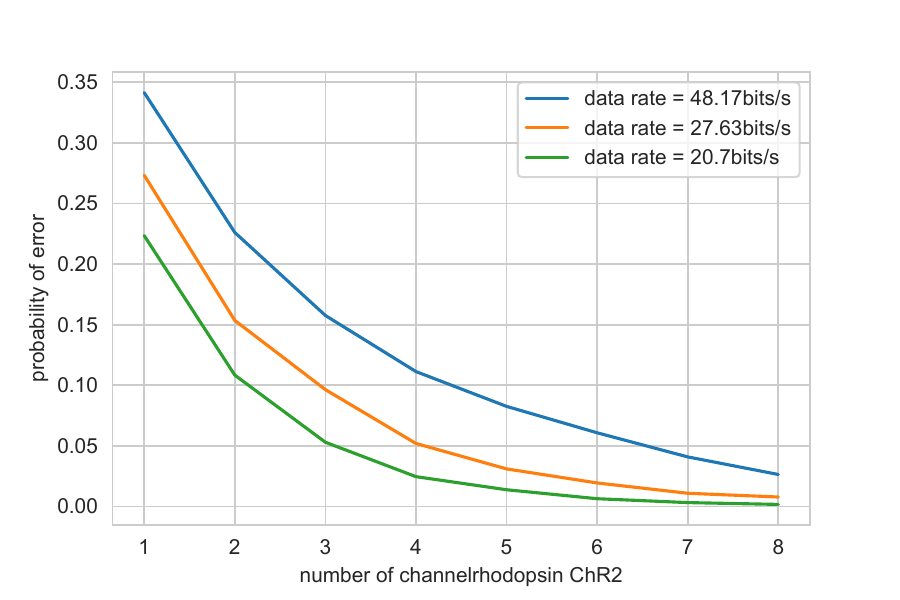}
\caption{Probability of error vs multiple ChR2 receptors when the data rate is fixed. As the number of ChR2 increases higher data rates can be achieved with a low probability of error. }
\label{figfixedrate}
\end{figure}

The effect of using multiple receptors is further illustrated in Fig. \ref{figcompdandp} where the probability of error is plotted against the data rate showing that the probability of error keeps on decreasing as the number of receptors increases. This  can be explained further in Fig.  \ref{figfixedrate}, where we make a plot of error probability with respect to the number of ChR2  for fixed data rates. The curves show that the probability of error is higher for high data rates however, as the number of channelrhodopsin increases, it becomes lower.\par
It may be possible to receive multiple channelrhodopsin receptors to obtain higher data rates, for example using a molecular MIMO technique \cite{channelrhodopsin2018}. Moreover, while the probability of error is in some cases relatively high, error-correcting codes can be used to increase reliability.


\subsection{Probability of error in noise}
\begin{figure}[h!]
\centering
\includegraphics[width=\columnwidth]{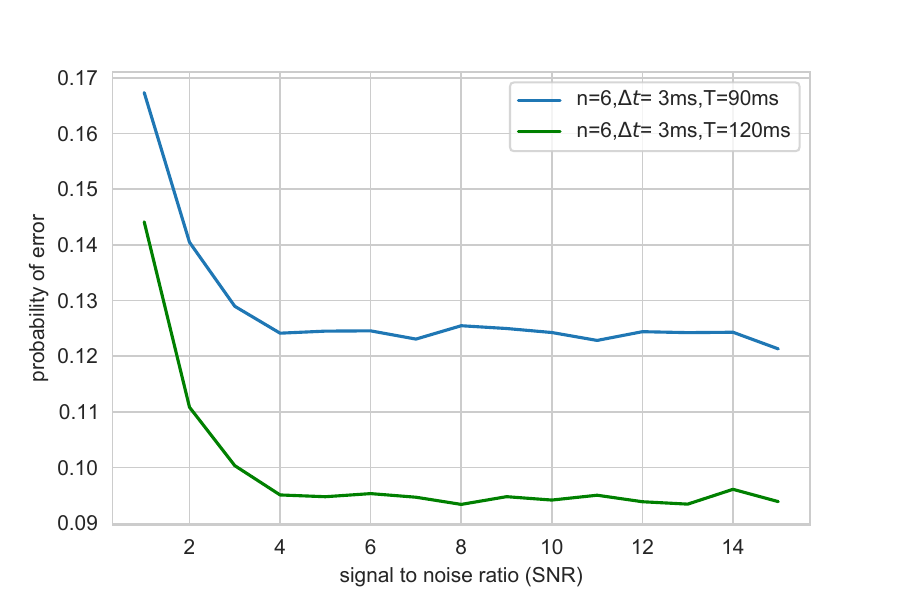}
\caption{Probability of error vs signal to noise ratio (SNR) for $n=6$ and when total time $T$ is different. The probability of error decreases with SNR.}
\label{snr6}
\end{figure}

In subsection III F, a light source is supposed to emit a fluctuating number of photons in a given time interval. Fig. \ref{snr6} shows that the probability of error decreases with the increase of signal-to-noise ratio (SNR) when observing states are six and $\Delta t$ = $3$ ms. The probability of errors decreases significantly as the total time T increases. 
As per (\ref{eqn:optical noise}), SNR is directly proportional to the square root of the average number of photons in a given interval $\lambda$. Therefore an increase in SNR is attributed to an increase in $\lambda$ that, in turn decreases the probability of error. 
The effect of SNR can be further explored in Fig. \ref{figsnr6,3}
\begin{figure}[h!]
\centering
\includegraphics[width=\columnwidth]{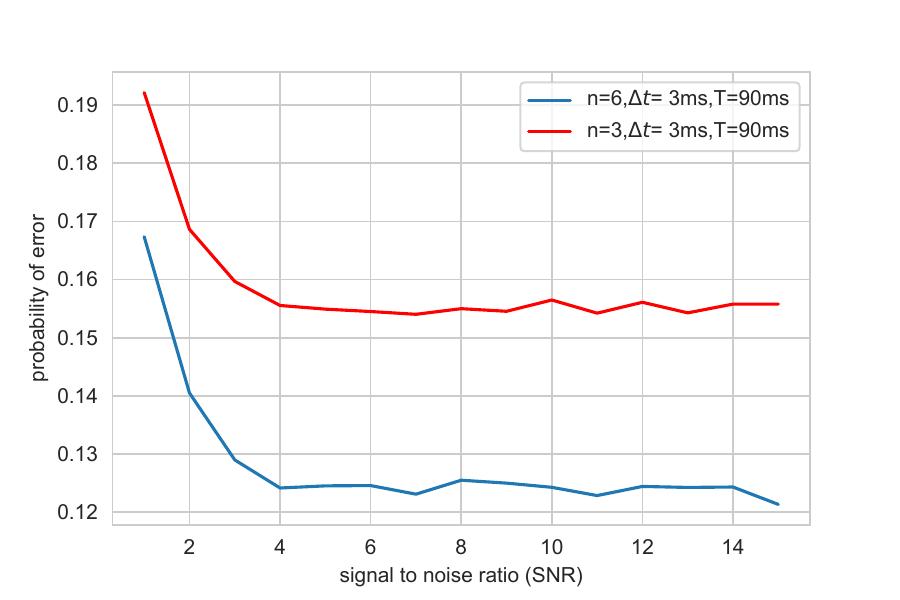}
\caption{Comparison of the probability of error vs signal to noise ratio (SNR) for $n=6$ and $n=3$ when $\Delta t$ is same.}
\label{figsnr6,3}
\end{figure}
the effect of SNR is more on three state model rather than on six states. Fig. \ref{figsnr6,3} shows the comparison of $n=3$ and $n=6$ in terms of signal-to-noise ratio when $\Delta t=3ms$.\par
\begin{figure}[h!]
\centering
\includegraphics[width=\columnwidth]{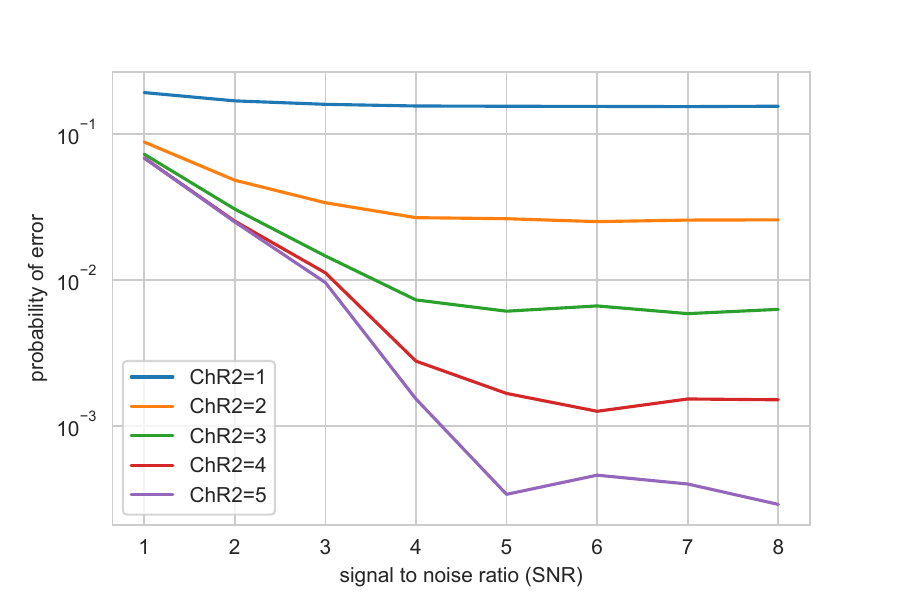}
\caption{Comparison of the probability of error vs signal to noise ratio (SNR) for different numbers of channelrhodopsins.}
\label{figmultiple}
\end{figure}

Similarly, we analyze the effect of SNR on multiple channelrhodopsins. As shown in Fig. \ref{figmultiple} where the number of one ChR2 is compared with two, three, four and five ChR2 for fixed $\Delta t$., the number of channelrhodopsin increases, and the probability of error decreases because the uncertainty is caused by the random opening and closing time of ChR2. We can see in Fig. \ref{snr6}, 
and Fig. \ref{figmultiple} that curves are flattened out as the signal-to-noise ratio increases. The rise in signal-to-noise ratio is linked to a higher light intensity level. The probability of error decreases as light intensity increases to a certain threshold; beyond this point, it stabilizes. This phenomenon may be attributed to the idea that lower light intensities might not be adequate to activate ChR2 effectively, thereby leading to a higher likelihood of errors. This is already illustrated in Fig. \ref{fig8} where an increase in light intensity level contributed to the probability of error up to a certain point.

 Based on the results, it is clear that the uncertainty is linked to the number of ChR2, as long as there are enough photons to activate them.

\section{Conclusion}
This work has presented the possibility of communication using a single light-sensitive bio-receptor ChR2. The concept of our proposed system is demonstrated by simple model, consisting of transmitter (light source) and receiver (ChR2) connected by a free space, with information expressed by the illumination of a light source. Data rates achieved by this simple method show that the proposed scheme can be considered a potential candidate for a future biomolecular communication system. \rev{ Furthermore, the photo-acoustic property of ChR2 \cite{walter2023time} renders it suitable for acoustic communications, enabling the transfer of information within nanoscale networks \cite{santagati2013opto}. In addition to its change of state, ChR2 undergoes volumetric changes compared to its parental state. This characteristic can be leveraged to transmit information from body-area nano-networks to the outside world, providing access to healthcare providers.}
In future work, it would be useful to consider multiple channelrhodopsin receptors to improve the information rate, while information-theoretic analysis of the channel and communication system may be performed to optimize the information rates.


\bibliographystyle{ieeetr}
\bibliography{globecom21}
\begin{IEEEbiography}[{\includegraphics[width=1in,height=1.25in,clip]{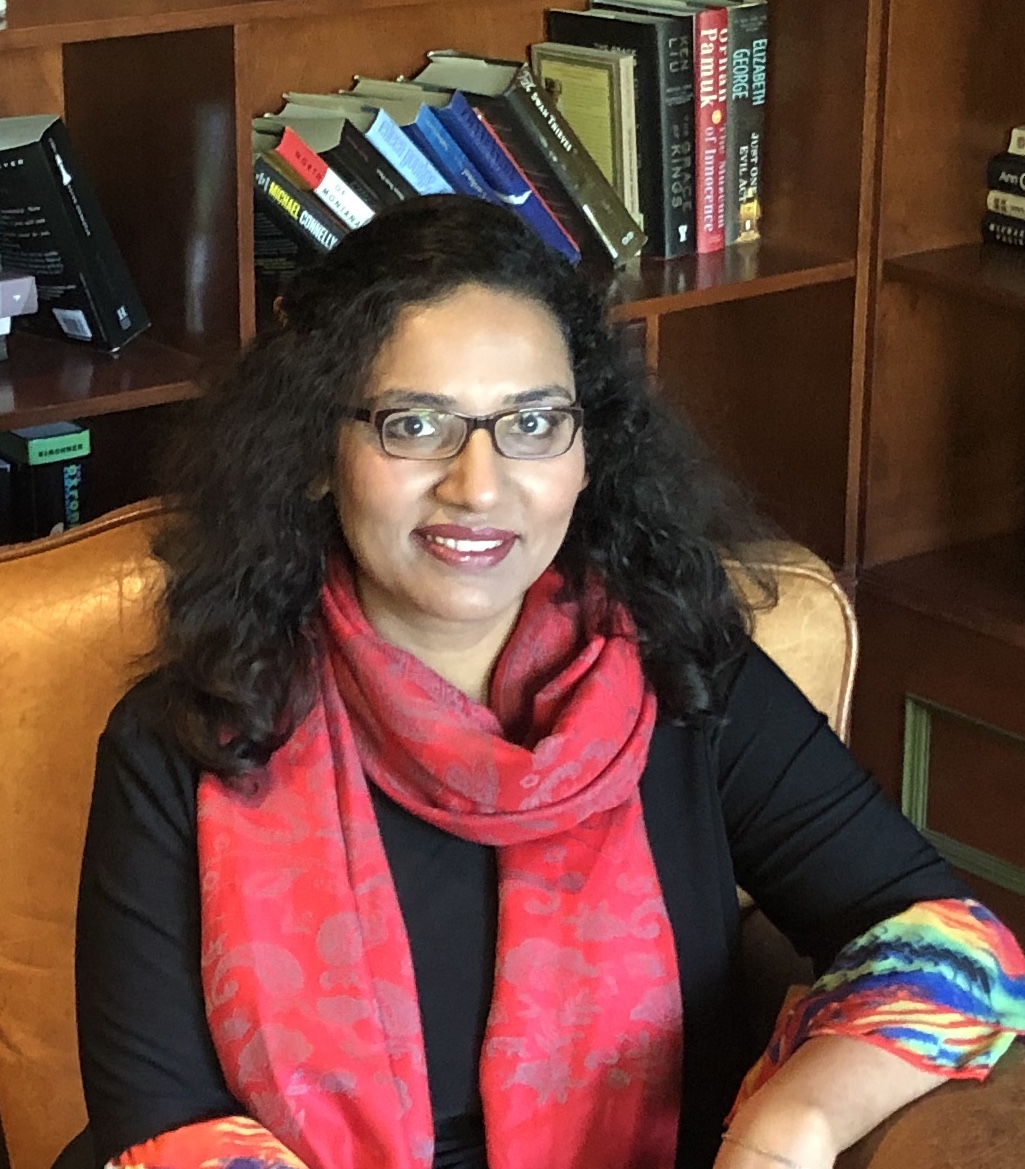}}]{Taha Sajjad} (Student Member, IEEE) is a graduate (Ph.D.) student in Lassonde Electrical and Computer Science
department York University. She is currently working in Eckford’s Lab on the application of
molecular communication. Taha did her M.A.Sc. in electrical engineering from
University of Engineering and Technology (UET), Pakistan, in 2007. Her research has revolved
around multidisciplinary areas, such as communications, embedded systems and biomedical
instrumentation. Taha has also translated a comprehensive book on mind sciences, which
explains how mind power could change one’s life. Moreover, she has been involved in various volunteer activities in the graduate student association and Canadian Science and Policy Centre(CSPC).
\end{IEEEbiography}

\begin{IEEEbiography}[{\includegraphics[width=1in,height=1.25in,clip]{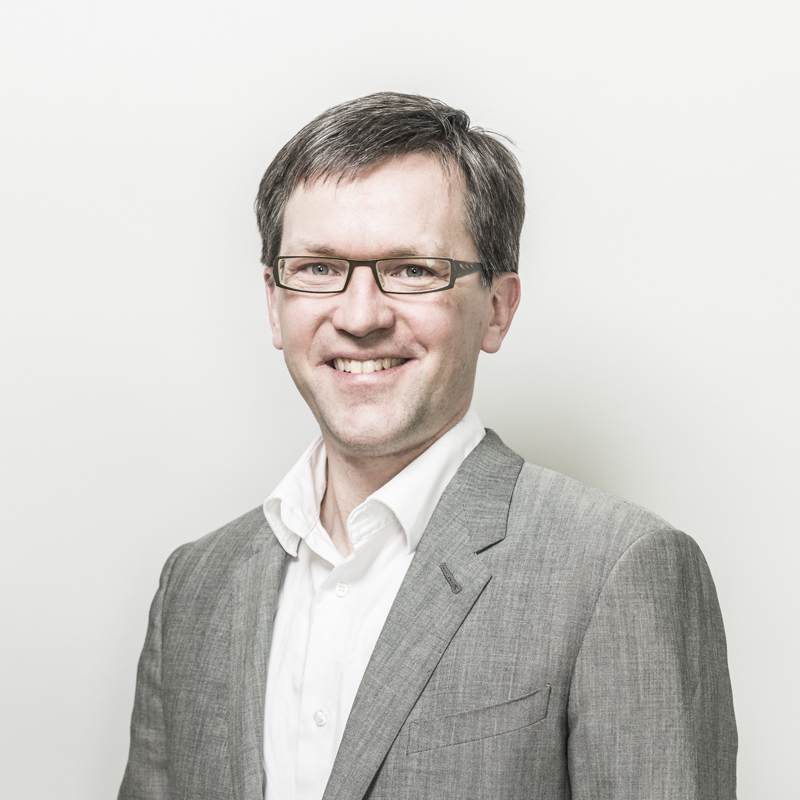}}]{Andrew W. Eckford} (Senior Member, IEEE) received the B.Eng. degree in electrical engineering from the Royal Military College of Canada in 1996, and the M.A.Sc. and Ph.D. degrees in electrical engineering from the University of Toronto in 1999 and 2004, respectively. He was a Postdoctoral Fellowship with the University of Notre Dame and the University of Toronto, prior to taking up a faculty position with York, in 2006. He is an Associate Professor with the Department of Electrical Engineering and Computer Science, York University, Toronto, ON, Canada. He has held courtesy appointments with the University of Toronto and Case Western Reserve University. In 2018, he was named a Senior Fellow of Massey College, Toronto. He is also a coauthor of the textbook Molecular Communication (Cambridge University Press). His research interests include the application of information theory to biology and the design of communication systems using molecular and biological techniques. His research has been covered in media, including The Economist, The Wall Street Journal, and IEEE Spectrum. His research received the 2015 IET Communications Innovation Award, and was a Finalist for the 2014 Bell Labs Prize.
\end{IEEEbiography}

\end{document}